\documentstyle[12pt]{article}
\nopagebreak

\def\3j#1#2#3#4#5#6{\mbox{$\left(\begin{array}{ccc}
#1 & #2 & #3 \\
#4 & #5 & #6
\end{array}
\right)$}}

\def\6j#1#2#3#4#5#6{\mbox{$\left\{\begin{array}{ccc}
#1 & #2 & #3 \\
#4 & #5 & #6
\end{array}
\right\}$}}

\def\eq{\begin{equation}}

\def\ee{\end{equation}}

\def\eqa{\begin{eqnarray}}

\def\eea{\end{eqnarray}}

\begin{document}


\centerline{\Large{\bf Density dependence of resonance broadening}}

\medskip

\centerline{\Large{\bf and shadowing effects in nuclear photoabsorption}}

\vskip 1.5cm

\centerline{\large{S.~Boffi$^{1,2}$, Ye. Golubeva$^3$, L. A. Kondratyuk$^4$, 
M. I. Krivoruchenko$^4$ and E. Perazzi$^1$}}

\vskip 1.0cm

\centerline{\small $^1$~Dipartimento di Fisica Nucleare e Teorica, 
Universit\`a di Pavia, Italy}

\centerline{\small $^2$~Istituto Nazionale di Fisica Nucleare, 
Sezione di Pavia, Pavia, Italy}

\centerline{\small $^3$~Institute for Nuclear Research of the Russian 
Academy of Sciences, Moscow, Russia}

\centerline{\small $^4$~Institute of Theoretical and Experimental Physics,
Moscow, Russia}

\vskip 1.5cm


\begin{abstract}

\noindent 

\end{abstract}

Medium effects as a function of the mass number $A$ are studied in the total
photonuclear cross section from the $\Delta$-resonance region up to the region
where shadowing effects are known to exist. A consistent picture is obtained
by simply assuming a density dependence of the different mechanisms of
resonance broadening and shadowing. The $\Delta$-mass shift is found
to increase with $A$.

\bigskip

PACS numbers: 25.20.-x

{\sl Keywords\/}: total photonuclear cross section, shadowing effects,
resonance broadening

\vskip 1.5cm

\clearpage


The total photoabsorption cross section on nuclei has been recently
measured at Frascati (see ref.~\cite{[Frascati],[Frascatibis]} and
references therein) and Mainz~\cite{[Mainz]} over the energy range 200-1200
MeV. A rather universal result is obtained for a variety of nuclei
indicating a basically incoherent volume photoabsorption mechanism. In the
region up to  500 MeV, the cross section per nucleon is not much affected by
medium  effects; the $\Delta$ peak due to the $P_{33}(1232)$ resonance is
clearly evident and not much distorted in comparison with the free nucleon.
In the region above 500 MeV, on the contrary, a strong damping and
flattening of the cross section are observed so that the $D_{13}$(1520) and
the $F_{15}$(1680) resonances, which are known to give the major
contribution to the second and third peak of the free nucleon cross section,
completely disappear even in light nuclei. Possible mechanisms responsible
for such a distortion of the baryon resonances in the nuclear medium have
been discussed~\cite{[Mauro]}-\cite{[Leonid]}, the major candidates being the
Fermi motion, the Pauli blocking and the collision broadening due to the
propagation of resonances inside nuclear matter. 

Beyond 2 GeV the observed reduction of the absolute value of the nuclear
photoabsorption cross section with respect to the free nucleon case might be
related to the onset of shadowing effects~\cite{[Weise]} which are known to
exist at higher energies on the basis of rather old data~\cite{[Bauer]}.
Consistency of resonance broadening and shadowing effects in the nuclear
photoabsorption process in the whole range of energies from the $\Delta$
peak up to the shadowing region has been recently shown within a generalized
photonuclear sum-rule approach~\cite{[BGKK]}. In this letter we focus on the
density dependence of the nuclear photoabsorption process. 

Following the approach of ref.~\cite{[Leonid]}, in order to isolate the
medium effects in the resonance region, the resonance mass $M_0$ and 
width $\Gamma_0$ for the free case have been taken from the fit of the
nucleon  photoabsorption cross section. Proton and neutron cross sections
have been considered separately as a sum of a smooth background  plus eight
Breit-Wigner resonances ($P_{33}(1232)$, $P_{11}(1440)$, $D_{13}(1520)$,
$S_{11}(1535)$, $D_{15}(1675)$, $F_{15}(1680)$, $D_{33}(1700)$,
$F_{37}(1950)$), which are believed to give the main contribution to the
total $\gamma N$ cross section below 1.2 GeV. The total nuclear
photoabsorption cross section is given in terms of the imaginary part of the
forward Compton scattering amplitude $\overline{\cal I}(\omega,\theta=0)$ as 

\eq
\frac{\sigma_{\gamma A}}{A}= -\frac{1}{2m\omega}\, {\rm Im}\, 
\overline{\cal I}(\omega,\theta=0),
\ee
where $\omega$ is the photon energy, $m$ is the nucleon mass and 
$\overline{\cal I} = \sum_{i}\, [(Z/A)\, \overline{\cal I}_{i,\gamma p} + 
(N/A)\, \overline{\cal I}_{i,\gamma n}]$. The elementary amplitude is given
by a convolution over the nucleon momentum distribution 
$\vert\phi({\vec p})\vert^{2}$ as

\eq
\overline{\cal I}_{\gamma N}(\omega,\theta=0)
= \int \frac{{\rm d}{\vec p}}{(2\pi)^{3}}\ \frac{\vert \phi({\vec
p})\vert^{2} g^{2}_{\gamma NN^{*}}}{[({\vec p}+{\vec k})^{2} - M^{2}
+ {\rm i}M\Gamma]},
\ee  
where $k = (\omega,{\vec k})$ and $p=(E,{\vec p})$ are the photon and
nucleon four-momenta, and $g_{\gamma NN^{*}}$ is the vertex function for
the $\gamma N\to N^*$ process. The effects of the medium are contained in
the mass shift $\delta M$ $(M = M_0 +\delta M)$ and in the width $\Gamma$,

\eq
\Gamma=\ \frac{\Gamma_{0}B_{P}+\ \Gamma^{*}}{S_{F}},\ee 
where $B_{P}$ is the Pauli blocking factor, $S_{F}$ the Fermi suppression
factor and $\Gamma^{*}$ the broadening due to collisions and interactions.
The Pauli blocking factor takes into account the increased lifetime of the
resonance due to the other nucleons occupying the momentum space below the
Fermi momentum $p_F$. The Fermi suppression factor $S_{F}$ is a decreasing
function of $x=2\omega p_{F}/M\Gamma_{0}$. As $x$ increases with $\omega$,
it is evident that the $\Delta(1232)$ is less affected by Fermi motion than
the other resonances. $\Gamma^{*}$ and $\delta M$ measure the broadening of
the resonance due to the interaction with surrounding nucleons; in the
optical pseudopotential approach, $\delta M$ and $\Gamma^{*}$ are
proportional to the real and imaginary part of the forward $NN^{*}$
scattering amplitude $f_{NN^{*}}(0)$, respectively.

A similar fit to the photofission cross section has already been carried out
for $^{238}$U in ref.~\cite{[Leonid]}, where the total number of
parameters were reduced to 5 by assuming $\delta M=0$ for all resonances
except for the $\Delta$ and a fixed broadening $\Gamma^{*}$ of
$P_{11}(1440)$, $S_{11}(1535)$, $D_{15}(1675)$, $F_{37}(1950)$. In fact these
parameters do not contribute significantly to the $\chi^{2}$.

In order to investigate medium effects we have introduced a density
dependence in all these parameters adjusting the number $A$ of nucleons, the
nuclear density $\rho$ and the Fermi momentum to the individual target
nuclei. Representative nuclei such as $^{12}$C, $^{63}$Cu and $^{208}$Pb were
considered.

Using the optical theorem, the broadening $\Gamma^{*}$ can be written as: 

\eq 
\Gamma^{*}= \rho\sigma^{*} v \gamma,
\ee
where $\sigma^{*}$ is the total cross section for the $NN^*$ interaction,
$\gamma$ is the Lorentz factor and $v$  the propagation velocity
($\sigma^{*}v$= 40 mb). The product $l=v\gamma$ gives the mean free path of
the resonance inside the nucleus. A constant $\rho=\rho_0$ has been assumed,
where $\rho_0$ is the normalization constant of the usual Woods-Saxon form
for $\rho(r)$.

The Fermi suppression factor and the Pauli blocking factor also depend 
on the nucleus through $p_{F}$, which is 0.221 GeV for $^{12}$C, 0.240 GeV
for $^{64}$Cu and 0.265 GeV  for $^{208}$Pb.

Finally, some dependence of the mass shift $\delta M$ of the $\Delta$ on the
number of nucleons has to be expected, as already found in electron 
scattering~\cite{[Sealock]}.

The results are shown in Fig. 1. While heavier nuclei keep no trace of the
second and third peak of the free-nucleon cross section, in the case of
$^{12}$C it is possible to see a little broad bump corresponding to the
second free-nucleon peak in both the theoretical curve and the data. On the
other hand, the $\Delta$ peak is well pronounced and the less affected by
collision broadening. However, it is clear that also the $\Delta$ peak is
damped for heavier nuclei, as an effect due to the medium density. In
agreement with data, the area under the resonance curve is unchanged, but
the width increases with $A$. For $^{12}$C we have $\Gamma^{*} = 51$ MeV for
the $\Delta$, while it is between 218 MeV and 235 MeV for the other
resonances ($D_{13}(1520)$, $F_{15}(1680)$, $D_{33}(1700)$). Similarly,
$\Gamma^{*} = 67 $ MeV for the $\Delta$ and between 280 MeV and 305 MeV for
the other resonances in the case of $^{63}$Cu and $\Gamma^{*} = 74$ MeV for
the $\Delta$ and between 315 MeV and 340 MeV for the others in the case of
$^{208}$Pb. Moreover, the $\Delta$-mass shift $\delta M$ shows a slight
dependence on $A$. Data on $^{12}$C are well described by $\delta M = 12$ MeV,
while $\delta M = 15$ MeV for $^{63}$Cu and $\delta M = 18$ MeV for
$^{208}Pb$. These values are comparable with those extracted in
ref.~\cite{[Frascatibis]} from experimental data, which can
be interpolated by a curve linearly dependent on A.   

Anyway, it is possible to say that $\sigma_{\gamma A}/A$ does not
strongly depend on A in the resonance region. On the contrary, in the
range of energies between 2 GeV and 20 GeV, shadowing effects have been
observed in the photonuclear cross section. These effects introduce a
correction to the cross section which can be expressed in terms of an
effective number of nucleons, $A_{\rm eff}$, as 

\eq
\sigma_{\gamma A}(\omega) = A\sigma_{\gamma N}-\Delta\sigma =
A_{\rm eff}(\omega)\sigma_{\gamma N} .
\ee

As discussed in many papers (see, e.g., \cite{[Weise],[Bauer],[Gribov]}), in
the region above 1 GeV diffractively produced intermediate hadronic states
occur by analogy with the case of hadron scattering by nuclei. The main
contribution comes from the lightest vector mesons, $\rho$, $\omega$ and
$\phi$ (the $\rho$-production threshold is 1.08 GeV).

In the eikonal approximation the Green function for the vector meson $V$
propagating from ${\vec r} = ({\vec b},z)$ to ${\vec r\,}' = ({\vec
b\,}',z')$ with momenutm  ${\vec k}_V$ can be written in the
form~\cite{[BGKK]}

\eq
G_V({\vec b\,}',z';{\vec b}, z) = \frac{1}{2{\rm i}k_V} {\rm exp}
\left\{{\rm i} \int_{z}^{z'}{\rm d}\xi\left[ k_V + \frac{2\pi}{k_V}f_V(0)
\rho({\vec b},\xi)\right]\right\}\,\delta({\vec b}-{\vec b\,}')\,
\theta(z-z') ,
\label{eq:green}
\ee
where ${\vec b}$ is the impact parameter, $f_V(0)$ is the forward $VN$
scattering amplitude  and the $z$ axis is taken along the 
resonance momentum ${\vec k}$.

Using this Green function we can easily find the shadowing correction as

\eq
\Delta\sigma = \frac{4\pi}{k}\, {\rm Im}\, F^{(2)}_{\gamma\gamma}, 
\label{eq:delta}
\ee
where

\eq
F^{(2)}_{\gamma\gamma} = -4\pi \sum_V 
\int{\rm d}{\vec r}\int{\rm d}{\vec r\,}'
\,{\rm e}^{-{\rm i}{\vec k}\cdot{\vec r\,}'} 
f_{\gamma V}(0) \rho_2({\vec r},{\vec r\,}') G_V({\vec r\,}',{\vec r})\, 
f_{V\gamma}(0)\,{\rm e}^{{\rm i}{\vec k}\cdot{\vec r}} ,
\label{eq:effdue}
\ee
with $f_{\gamma V}(0)$ being the forward $\gamma V$ transition amplitude.
In eq.~(\ref{eq:effdue}) $\rho({\vec r},{\vec r\,}')$ is the two-body density
function. In the case of an uncorrelated system, the usual approximation of
$\rho_2$ in terms of one-body densities, i.e. $\rho_2({\vec r},{\vec r\,}') =
\rho({\vec r})\rho({\vec r\,}')$ makes the above formulae to be equivalent to
the optical-model approximation to the multiple scattering expansion used in
ref.~\cite{[Bauer]}.

We incorporate correlations between pairs of nucleons choosing the simple
Bessel function parametrization for the two-body correlation function

\eq
\Delta({\vec r},{\vec r\,}') = \rho_2({\vec r},{\vec r\,}') - 
\rho({\vec r})\rho({\vec r\,}') = - j_0(q_c\vert{\vec r}-{\vec r\,}'\vert)
\rho({\vec r})\rho({\vec r}'),
\label{eq:corr}
\ee
where $q_c = 780$ MeV~\cite{[Piller]}.

The results shown in Fig. 2 have been obtained with the parametrization of
the $f_{\gamma V}$ amplitudes in model I of ref.~\cite{[Bauer]} putting
however the real parts of the forward scattering amplitudes $f_{\gamma V}(0)$
equal to 0. A rather satisfactory agreement with data is gained, especially
when including correlations whose contribution to $A_{\rm eff}/A$ is about
10\% for copper and lead and about 5\% for carbon. $A_{\rm eff}/A$ decreases
with energy towards an asymptotic constant value in agreement with the
finding of  ref.~\cite{[Weise]}, i.e.

\eq
A_{\rm eff}/A = A^{-0.09}.
\ee
According to our calculations this constant value is reached more rapidly
for lighter nuclei, already above 6 GeV for carbon. This is also the
trend of the available data. However, better data are certainly welcome to
test the density dependence of the shadowing effects.

Taking into account correlations, there is a region of energies (below 2 GeV)
where $A_{\rm eff}/A > 1$, i.e. antishadowing takes place. This was also
found in the case of $^{238}$U~\cite{[BGKK]} and happens because correlations
essentially change the dependence of the amplitude $F_{\gamma\gamma}^{(2)}$
on the longitudinal momentum transfer $q_L= k - k_V = \omega - \sqrt{\omega^2
- m_V^2}$ in the region where $q_L r > 1$. No data exist in this energy
region which is the typical range of energies explored at
CEBAF~\cite{[CEBAF]} and ELSA~\cite{[ELSA]}. Therefore we hope that new data
will help to understand the onset of shadowing.

\bigskip

This work was supported by INTAS grant no. 93-79. Two of us (Ye. G. and L. A.
K.) are grateful to Dipartimento di Fisica Nucleare e Teorica of the
University of Pavia for kind hospitality.


\bigskip


\centerline{\bf Figure captions}

\bigskip

Fig. 1. The total photoabsorption cross section for $^{12}$C, $^{63}$Cu
and $^{208}$Pb. Data from~\cite{[Frascatibis]}.

\bigskip

Fig. 2. Results of calculations of $A_{\rm eff}/A$ for $^{12}$C, $^{63}$Cu
and $^{208}$Pb taking into account contributions of $\rho$, $\omega$, and
$\phi$. The solid and dashed curves refer to calculations with and without
correlations. The data are from SLAC~\cite{[SLAC]} (circles) and
Cornell~\cite{[Cornell]} (triangles).


\begin{thebibliography}{99}

\bibitem{[Frascati]}
{N.~Bianchi et al., Phys. Lett. B325 (1994) 333.}

\bibitem{[Frascatibis]}
{N.~Bianchi et al., preprint LNF-95/053 (P).}

\bibitem{[Mainz]}
{Th. Frommhold et al., Phys. Lett. B295 (1992) 28.}

\bibitem{[Mauro]}
{M. M. Giannini and E. Santopinto, Phys. Rev. C49 (1994) R1258.}

\bibitem{[Wanda]}
{W. M. Alberico, G. Gervino, and A. Lavagno, Phys. Lett. B321 (1994) 177.}

\bibitem{[Leonid]}
{L. A. Kondratyuk et al., Nucl. Phys. A579 (1994) 453.}

\bibitem{[Weise]}
{W. Weise, Phys. Rev. Lett. 31 (1973) 773; Phys. Rep. 13 (1974) 53.}

\bibitem{[Bauer]}
{Th. Bauer et al., Rev. Mod. Phys. 50 (1978) 261.}

\bibitem{[BGKK]}
{S. Boffi, Ye. Golubeva, L. A. Kondratyuk and M. I. Krivoruchenko, preprint
FNT/T-96/5 and Nucl. Phys. to be published.}

\bibitem{[Sealock]}
{R. M. Sealock et al., Phys. Rev. Lett. 62 (1989) 1350.}

\bibitem{[Gribov]}
{V. N. Gribov, Zh. Eksp. Teor. Fiz. 57 (1969) 1306 [Sov. Phys. JETP 30 (1970)
709.}

\bibitem{[Piller]}
{G. Piller and W. Weise, Phys. Rev. 42 (1990) R1834.}

\bibitem{[SLAC]}
{Caldwell et al., Phys. Rev. D7 (1973) 1362.}

\bibitem{[Cornell]}
{Michalowski et al., Phys. Rev. Lett. 39 (1977) 737.}

\bibitem{[CEBAF]}
{B. L. Berman, N. Bianchi, and V. Muccifora, CEBAF approved experiment
E-93-019.}

\bibitem{[ELSA]}
{N. Bianchi et al., ELSA approved experiment 1996.}

\end{thebibliography}
\end{document}